\documentclass[]{spie} 
\usepackage[pdfborder= {0 0 0 }]{hyperref}
\usepackage[all]{hypcap}
\usepackage{amssymb}
\usepackage{graphicx}
\newcommand{\email}[1]{\href{mailto:#1}{\nolinkurl{#1}}}

\title{Schwarzschild-Couder Telescope for the Cherenkov Telescope Array}

\author{Kevin J. Meagher\supit{a}\footnote{\ E-mail address: \email{kevin.meagher@physics.gatech.edu}}\ \ for the Cherenkov Telescope Array Consortium
\skiplinehalf
\supit{a}School of Physics and Center for Relativistic Astrophysics, Georgia Institute of Technology, 837 State Street NW, Atlanta, GA 30332
}
\begin{document} 
\maketitle 

\begin{abstract}
The Cherenkov Telescope Array (CTA) is the next major ground-based observatory for gamma-ray astronomy. With CTA gamma-ray sources will be studied in the very-high energy gamma-ray range of a few tens of GeV to 100\,TeV with up to ten times better sensitivity than available with current generation instruments. We discuss the proposed US contribution to CTA that comprises imaging atmospheric Cherenkov telescope with Schwarzschild-Couder (SC) optics. Key features of the SC telescope are a wide field of view of eight degrees, a finely pixelated camera with silicon photomultipliers as photon detectors, and a compact and power efficient 1\,GS/s readout. The progress in both the optical system and camera development are discussed in this paper.
\end{abstract}



\section{INTRODUCTION}
\label{sec:intro}  

The very-high energy (VHE) gamma-ray band (10\,GeV to 100\,TeV) is a unique window in the electromagnetic spectrum. Imaging atmospheric Cherenkov telescopes (IACT) like VERITAS, HESS and MAGIC have provided us with stunning insight into some of the most extreme and energetic processes in the violent universe. The number of known VHE gamma-ray emitting sources has increased from about 10 to more than 140 in less than 10 years, which far exceeded even the most optimistic predictions. This unexpected boost in source count is one reason to believe that VHE gamma-ray emission is a common phenomenon rather than a rare exception. Our view of the VHE sky is clearly sensitivity limited, which is why a global effort is underway to design, construct, and operate the Cherenkov Telescope Array (CTA) that will deliver ten times higher sensitivity than currently operating Cherenkov telescopes \cite{2013APh....43....3A,2011ExA....32..193A}. The main science goals of CTA are to understand the origin of cosmic rays, to understand the nature of particle acceleration from black holes, and to investigate the enigmatic nature of dark matter. It is expected that CTA will detect more than 1000 new VHE sources.

Cherenkov telescopes collect Cherenkov photons from air showers of secondary particles that result when a VHE gamma-ray or charged cosmic ray interacts with the earth's atmosphere. The Cherenkov photons are projected onto cameras with about 0.1 degree angular resolution, and an image of the air shower is recorded. The sensitivity of present IACTs is limited by the size of the area that the Cherenkov light of one air shower illuminates on the ground (about 100,000\,m$^2$). CTA achieves a factor of ten improvement in sensitivity by building an array of tens of IACTs that covers an area 100 times larger than the Cherenkov light pool. An additional significant contribution to the sensitivity of CTA comes from the fact that the same shower is viewed by many more telescopes. This is where the CTA group in the United States contributes to CTA---namely by filling the baseline array of CTA with IACTs that improve the baseline sensitivity by a factor of about two between 100\,GeV and few TeV according to latest Monte Carlo simulation.

A design driver of CTA is to achieve ten times higher sensitivity while simultaneously optimizing costs. This is achieved by constructing IACTs of three different sizes: large sized telescopes (LST), medium sized telescopes (MST), and small sized telescopes (SST). The center of the array contains four LSTs that each have a 23\,m diameter light collection surface and are separated by about 100\,m. The larger mirror surface of an LST collects more Cherenkov light than a smaller telescope and thus is more sensitive at lower energies. Above a few hundred GeV, the smaller mirror surface of the MSTs does not limit the detection and reconstruction efficiency of gamma rays. The MSTs are placed on a grid with a spacing of 150\,m. The SSTs are designed to be fully efficient above a few TeV and surround the LST and MST telescopes with a spacing of 300\,m.

Full-sky coverage will be achieved with CTA by constructing two arrays, one in the Southern Hemisphere and one in the Northern Hemisphere. The southern array will focus on the study of galactic objects and will incorporates all three telescope sizes. The northern array will focus mostly on extra-galactic objects, which suffer from extinction of the highest energy gamma rays due to energy-dependent interaction of the gamma rays with the extragalactic background light and therefore does not include SSTs. The Northern observatory is planned to consist of 4 LSTs and 15 MSTs to extend over an area of nearly 1\,km$^2$. In the baseline plan, the Southern observatory will consist of 4 LSTs, 25 MSTs, and around 70 SSTs over an area of nearly 10\,km$^2$.

\begin{figure}
  \begin{center}
    \begin{tabular}{c}
      \includegraphics[width=0.5\columnwidth]{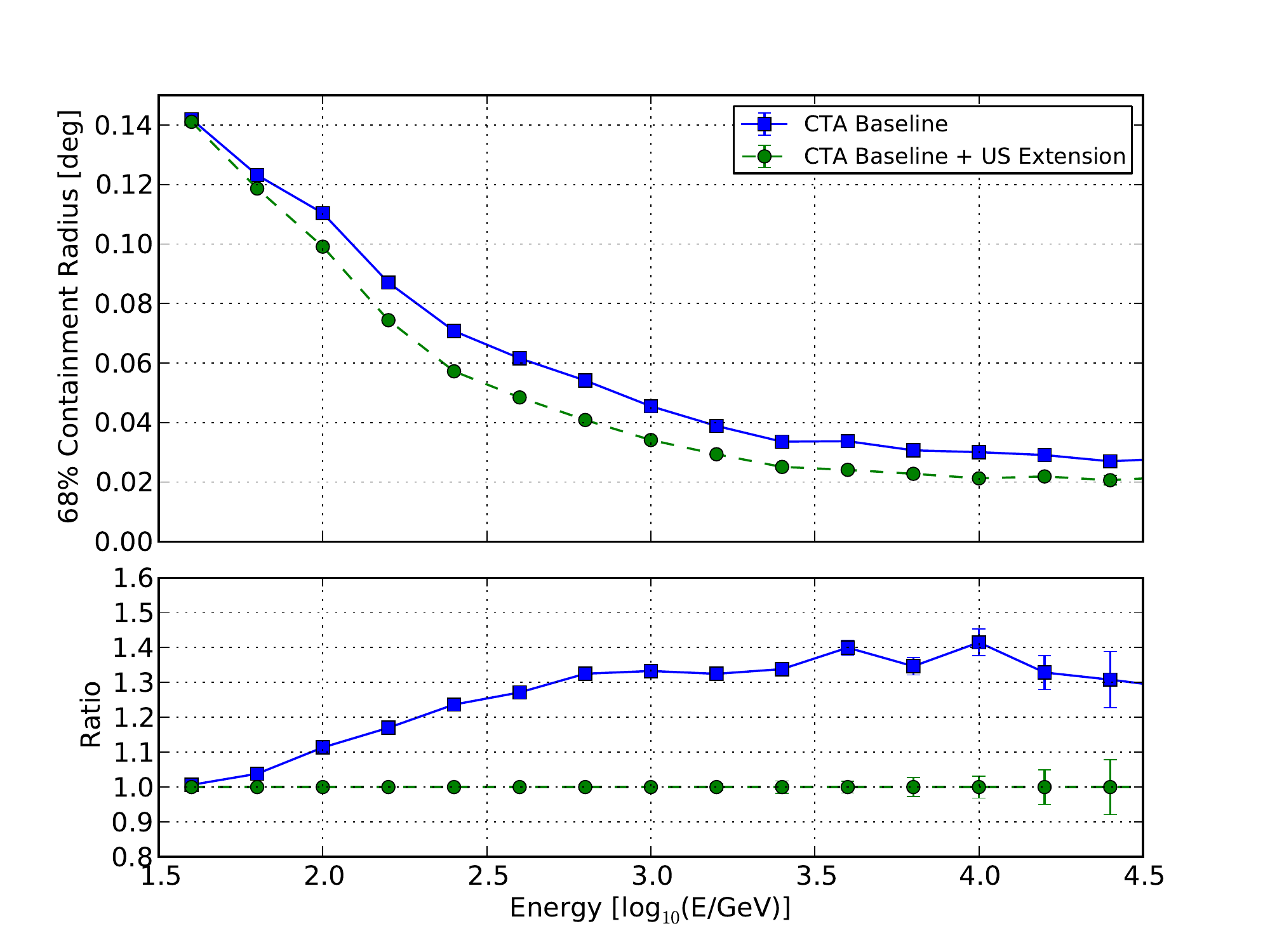}%
      \includegraphics[width=0.5\columnwidth]{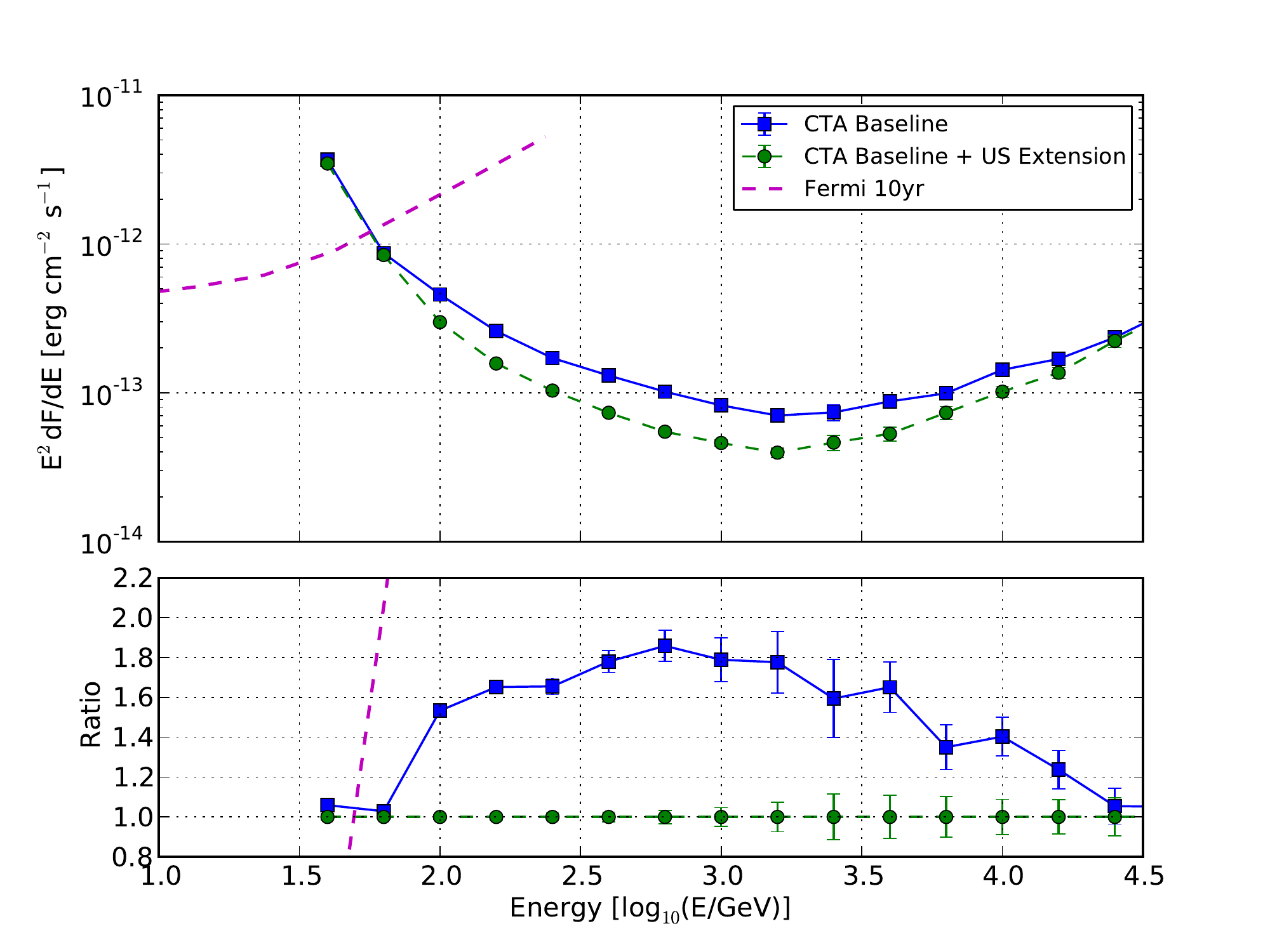}
    \end{tabular}
  \end{center}
  \caption[example]{\label{fig:simulation} 
    Simulations of CTA baseline and with the proposed US extension. Baseline refers to an array with 4 LSTs, 25 Davies-Cotton MSTs, and 37 SSTs\cite{2013arXiv1307.2773B}, and the US Extension refers to an additional 24 SCTs. \emph{Left}: 68\% containment radius of gamma rays from a point source as a function of energy both with the SCTs and without. The SCT contribution improves the angular resolution by 40\%. \emph{Right}: Differential sensitivity of the CTA baseline alone and with the SCT extension as a function of energy. The SCT extension doubles the sensitivity of CTA.
  }
\end{figure}

Two designs have been proposed for the MST: a traditional, single-mirror Davies-Cotton telescope for the baseline plan of CTA and an innovative, two-mirror Schwarzschild-Couder telescope as an infill of the baseline plan. The baseline plane consists of 25 MSTs, the CTA-US group has proposed to extend by contributing with 24 Schwarzschild-Couder telescopes (SCT). 
Figure \ref{fig:simulation} shows the factor-of-two change in sensitivity between the baseline design and the baseline with the SCT extension.

Funded by the National Science Foundation (NSF), the CTA-US group is presently constructing a prototype SCT (pSCT) to demonstrate the technical feasibility of the new design. The pSCT will be mounted on an unused foundation at the VERITAS site and integrated into the VERITAS array. VERITAS is an array of four 12-meter Davies-Cotton telescopes at the Fred L.\ Whipple observatory in Arizona \cite{2006APh....25..391H}.

In this paper I discuss the present status of the pSCT. While I intend to give a complete overview of the project, I put more emphasis on the development of the camera.

\section{Optics}
\label{sec:optics}
The principle of the SCT optics was first developed by Karl Schwarzschild in 1905. He proposed a two-mirror design for optical telescopes with the aim to simultaneously remove spherical and comatic aberrations inherent in reflection telescopes\cite{1905MiGoe..10....1S}.
He found an analytic, aplanatic solution that minimizes aberrations near the optical axis.
Andre Couder improved upon Schwarzschild's design by adding curvature to the focal plane, which eliminates astigmatisms.
To date no Schwarzschild-Couder telescope has been built, mainly due to difficulties in manufacturing non-spherical mirrors that meet the necessary specifications. However, recent advances in mirror manufacturing have made it feasible to realize an SCT.

The optics design we employ is based on extensive ray-tracing simulations\cite{2007APh....28...10V}. The design is realized with a primary reflective surface that is 9.5\,m in diameter and has a 4.4\,m diameter hole in the center. The primary mirror consists of 48 panels arranged in two rings. The secondary mirror has a diameter of 5.4\,m and is constructed out of 24 panels, again in two rings. Ray-tracing simulations have shown that with this design, the optical point-spread function is 4 arcminutes or less over the entire field of view.
\begin{figure}
  \begin{center}
    \begin{tabular}{c}
      \includegraphics[height=7cm]{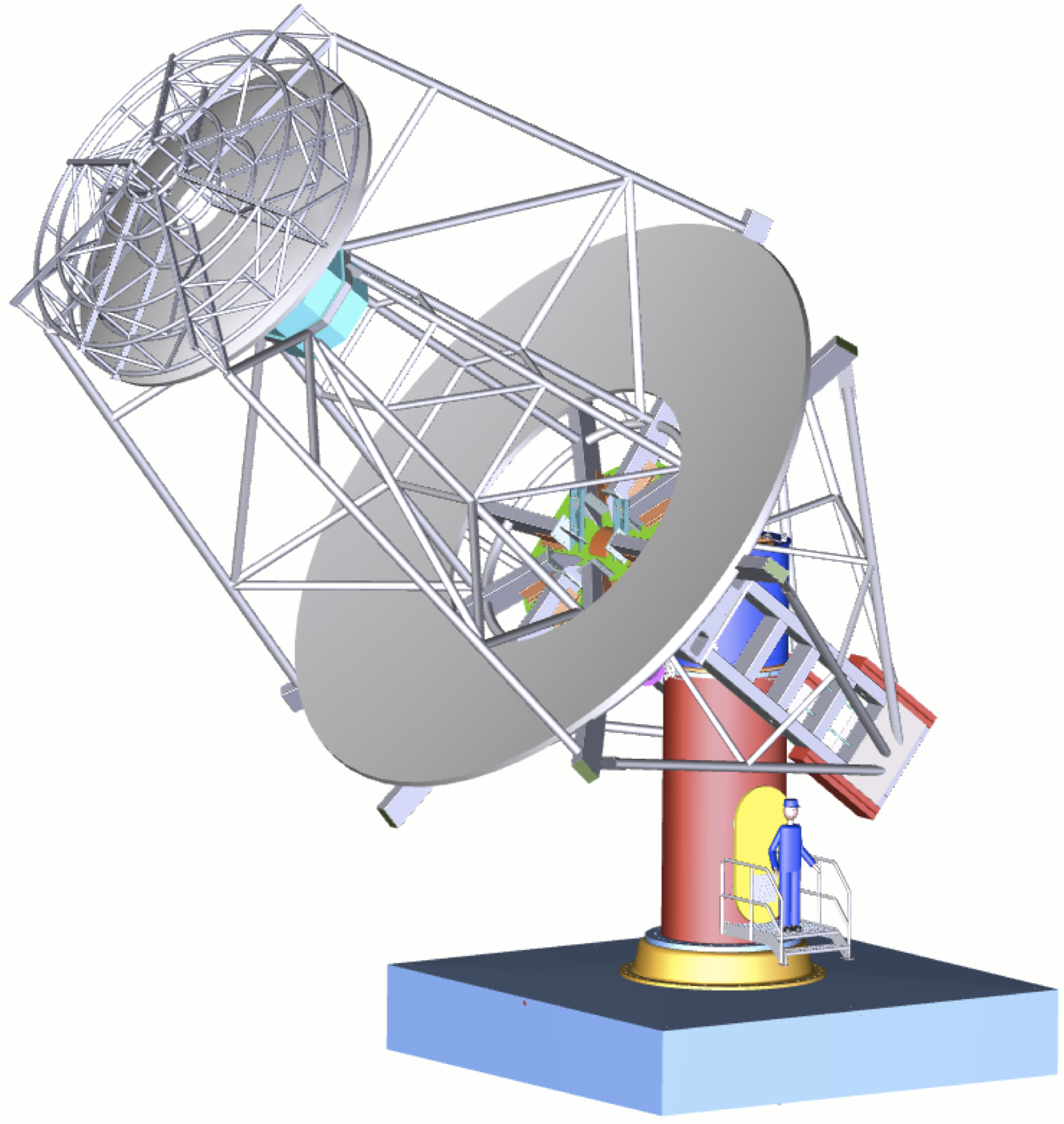}%
      \includegraphics[height=7cm]{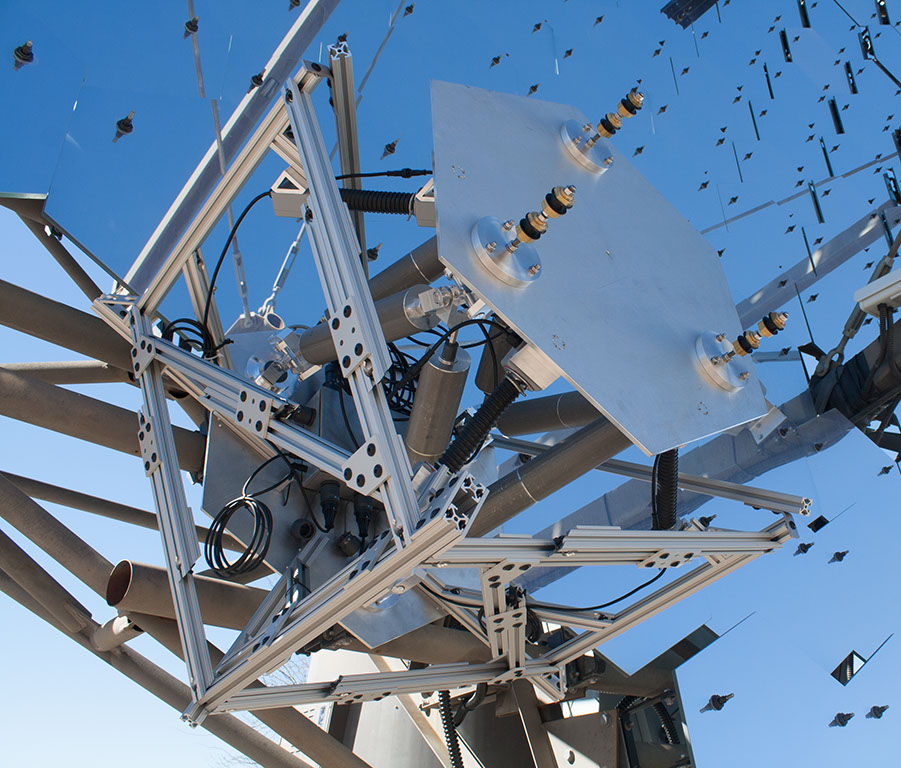}
    \end{tabular}
  \end{center}
  \caption[example]{\label{fig:telescope_and_hexapod} 
    The left panel shows a model of the SCT. The right panel shows a prototype of the hexapod mounted to one of the VERITAS telescopes.
  }
\end{figure}

Each mirror segment is mounted on six actuators in a Stewart Platform configuration for alignment and positioning. The actuators have a step size of 1.5\,$\mu$m over a range of 63\,mm. They are attached to a universal joint designed to allow five degrees of freedom while maintaining a hysteresis below several microns. The positioning of each segment relative to its neighbors is monitored by a series of “mirror panel edge sensors (MPES). Each sensor consists of a diode laser mounted on one mirror panel and aimed at a screen mounted on a neighboring panel and viewed by a commercial webcam. Three such sensors are mounted along the radial edges between mirror panels within each ring of panels. These “triads” are oriented to be mutually orthogonal, allowing their $X$ and $Y$ measurements together to constrain all six degrees of freedom between neighboring panels. Additional sensors are mounted between panels of the inner and outer rings of the primary and secondary. Laboratory tests have demonstrated that individual sensors have a resolution $< 5\,\mu\textrm{m}$. A field test including a prototype Stewart Platform and three MPES is underway at the VERITAS site; the setup for the field test is shown in Figure \ref{fig:telescope_and_hexapod}. 

The mirrors and optical support structure are mounted on an altitude-azimuth drive system (see Figure \ref{fig:telescope_and_hexapod}), which is the same system the baseline MST telescope uses. A mechanical prototype of the MST is operating in Adlershof near Berlin, Germany. Finite element analysis of the mechanical support structure have shown that the focal point does not change by more than $100\,\mu$m due to temperature changes and pointing the telescope at different elevations.

\section{Camera}

The camera of the SCT at located in the focus of the secondary mirror in between the primary and the secondary mirrors. Linear stages mounted to the frame of the camera allow positioning the camera with a precision of $100\,\mu$m along all three axes, which is a necessary requirement to achieve the design goal for the optical point-spread function. The photon sensitive area (focal plane) of the camera has a diameter of only 0.8\,m and covers a field of view of $8^\circ$. For comparison, the camera of the MST has a diameter of $\sim2.5$\,m and covers the same field of view as the SCT. The compactness of the SCT camera is a consequence of the demagnifying optics of the SCT and allows for the construction of a cost-efficient camera. An additional advantage of the smaller camera is the possibility to use silicon photomultipliers (SiPMs). SiPMs are mechanically and electrically more robust than photomultiplier tubes that are typically used in IACTs. 

The SCT camera is composed of 11,328 pixels with each pixel having a physical size of $6\times6$\,mm$^2$ and an angular size of 0.067$^\circ$. The pixels are grouped into modules, each of which is composed of 64 pixels. A fully equipped camera consists of 177 modules, where groups of 25 modules are arranged in so-called sectors of which there are nine per camera. For the pSCT only the central sector will be equipped with modules. The entire camera consumes about 3\,kW of electrical power. The heat produced by the camera is removed by two arrays of fans mounted to opposite sides of the camera housing and operating in a push-pull configuration.

\begin{figure}[t]
  \begin{center}
    \begin{tabular}{c}
      \includegraphics[width=0.49\textwidth]{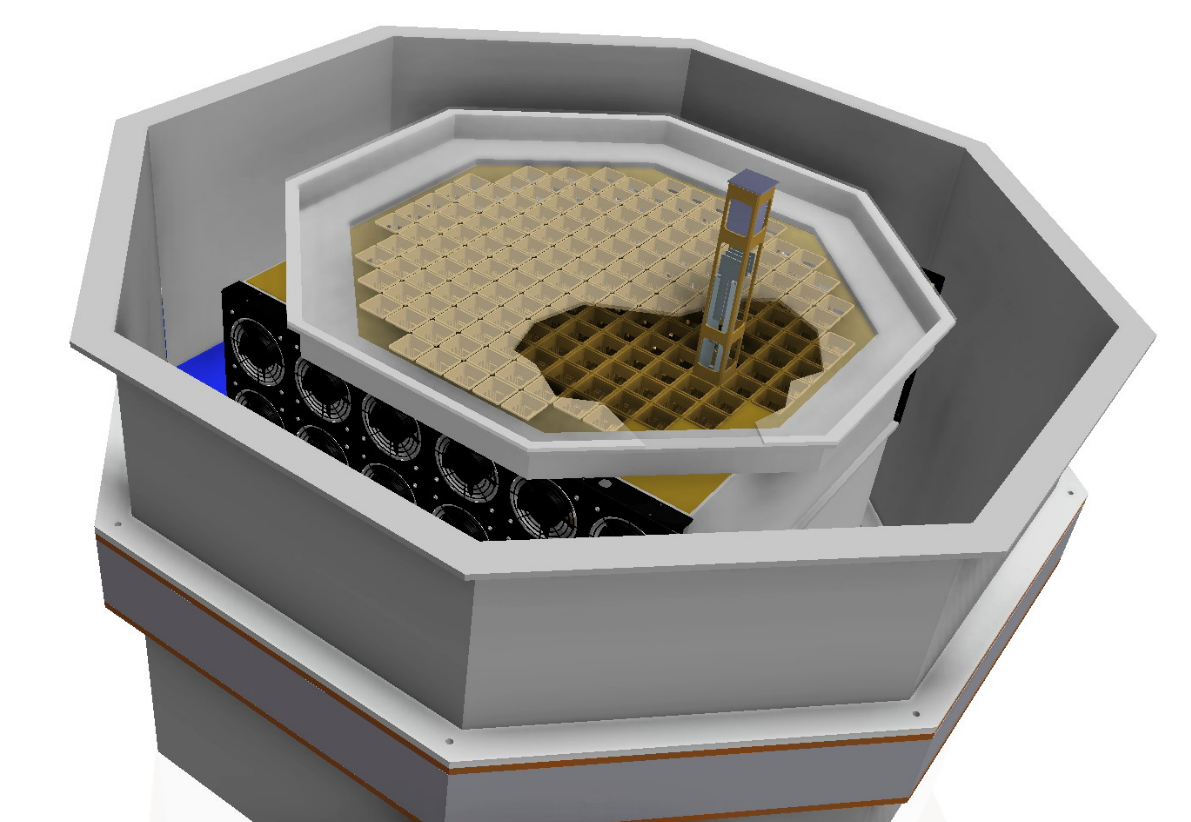}%
      \includegraphics[width=0.49\textwidth]{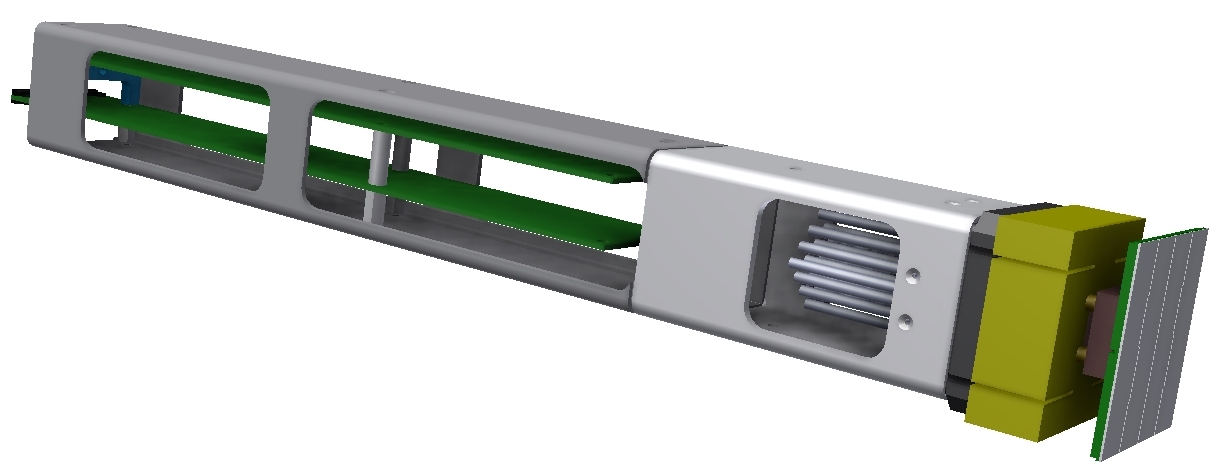}
    \end{tabular}
  \end{center}
  \caption{\label{fig:camera_and_module} 
   The left panel shows a drawing of the SCT camera. Modules that house the photon detectors and the readout electronics are inserted from the front into the camera. The right panel shows a more detailed view of one module.  
}
\end{figure}

A module not only hosts the photon detectors but also houses the front-end electronics and the signal digitization. Figure \ref{fig:camera_and_module} shows how a module is inserted from the front into the camera. The modules slide into the camera guided by carbon-fiber rods that exhibit negligible thermal expansion. Once inserted the top of a module rigidly rests on the front lattice of the camera, which provides the reference point for the absolute positioning of a module. Electrical contact with the backplane of each sector providing power and communications for the module is made through one multi-pin connector located in the back of each module.

For practical purposes a module is divided into the focal-plane module and the front-end electronics. The focal-plane module (FPM) is located in the front of the module and rests on the aluminum cage that gives the module its mechanical stability, and the front-end electronics (FEE) is mounted inside the cage.

\subsection{Focal Plane Module}
\label{sec:fp}
\begin{figure}[t]
  \begin{center}
    \begin{tabular}{c}
      \includegraphics[height=5cm]{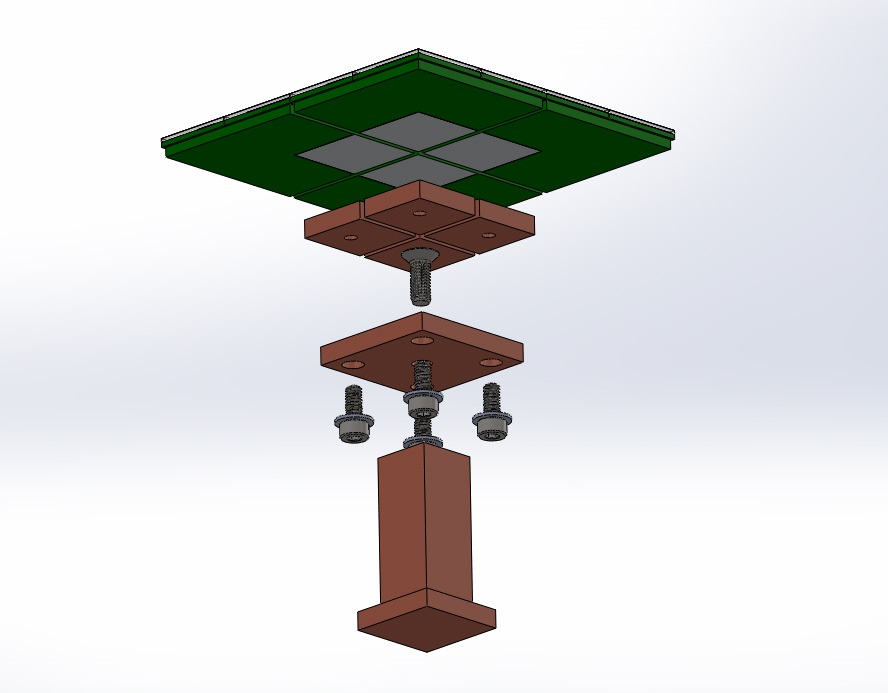}%
      \includegraphics[height=5cm]{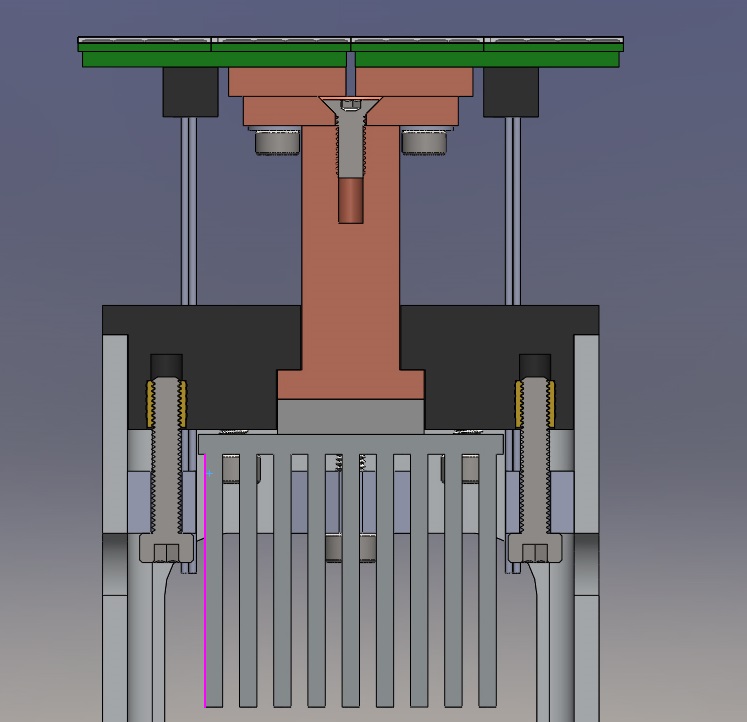}
    \end{tabular}
  \end{center}
  \caption[example]{\label{fig:focalplanemodule} 
Drawing of the components of the FPM and a cross section of a FPM mounted on a cage. The photon detectors are mounted on PCB (green). The four PCBs of a module are connected to a copper post that is cooled by a thermoelectric element attached to the bottom of the post.
}
\end{figure}

\enlargethispage{10mm}
Figure \ref{fig:focalplanemodule} shows an exploded view of the components of the FPM and a cross section of the FPM in its final position on top of a cage. The photon detectors are reflow soldered to four printed circuit boards (PCBs) called quadrants. Each quadrant is positioned along the optical axis approximating the ideal curved surface of the focal plane. Positioning is accomplished by adjusting the height of the central copper post and placing copper shims in between the square base that is mounted to the top of the post and the PCBs. In this scheme the vertical position can be adjusted to within $100\,\mu$m as has been verified in prototypes. 
\begin{figure}[t]
  \begin{center}
      \includegraphics[height=7cm]{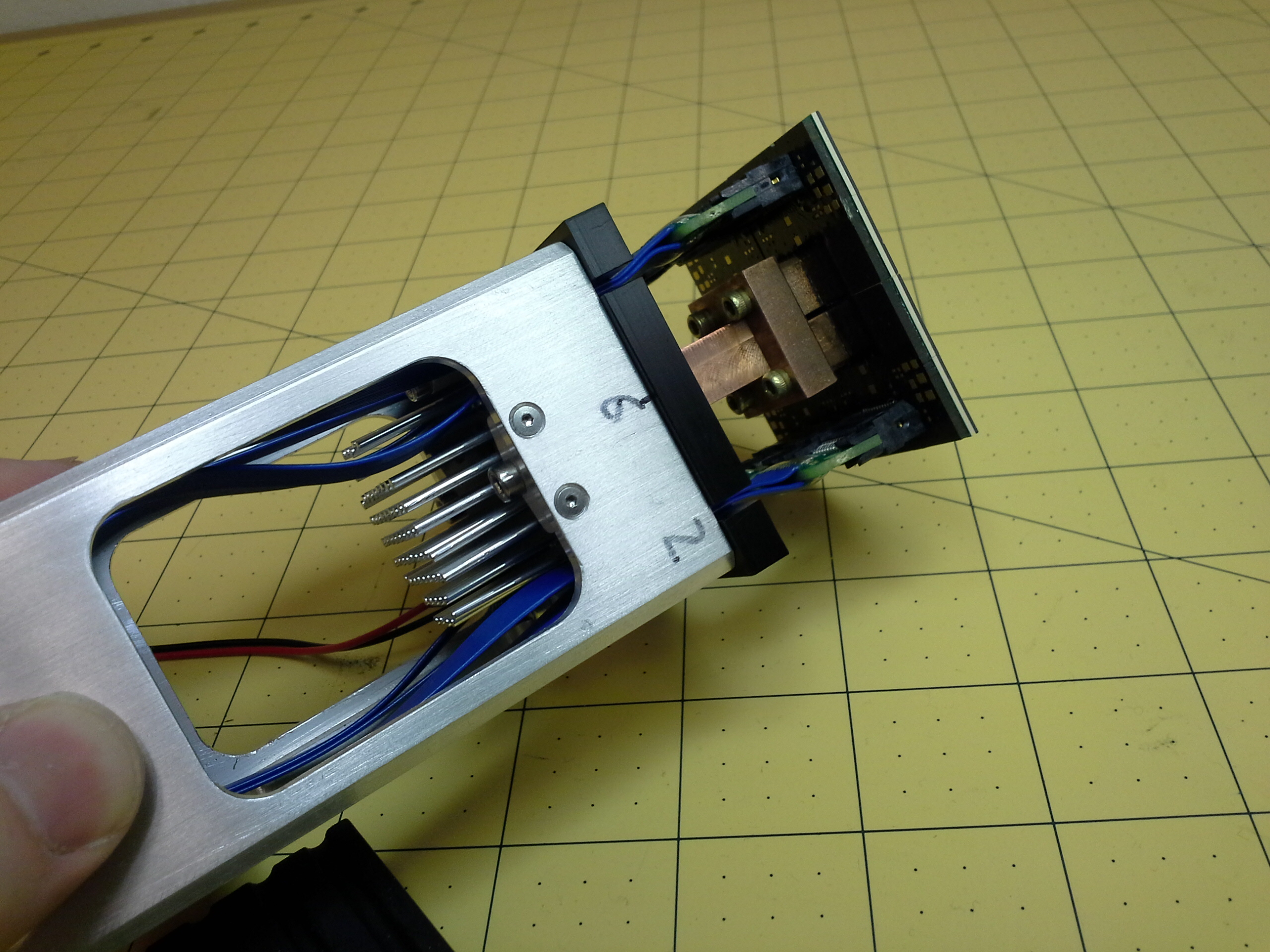}
  \end{center}
  \caption[example]{\label{fig:focalplanepicture} 
Prototype of the FPM mounted on the top of a module cage. 
  }
\end{figure}

Figure \ref{fig:focalplanepicture} shows the prototype of a FPM mounted on top of a cage. Electrical connections between the quadrants and the FEE are established by Samtec ECRD coaxial ribbon cable. 
\vspace{30mm}

\subsection{Photon detectors}
SiPM from Hamamatsu are used in the pSCT. The device, S12642-0404PA-50(X), is a tile with 16 SiPMs. Each SiPM in the tile is $3\times3$\,mm$^2$ in size and consists of 3600 avalanche diodes that operate independently in Geiger mode. The size of the avalanche cell is $50\times50$\,$\mu$m$^2$. On the PCB, the devices are reflow soldered on; four SiPMs are electrically connected in parallel to from one $6\times6$\,mm$^2$ pixel in the readout. 

\begin{figure}[t]
  \begin{center}
    \begin{tabular}{c}
      \includegraphics[height=7cm]{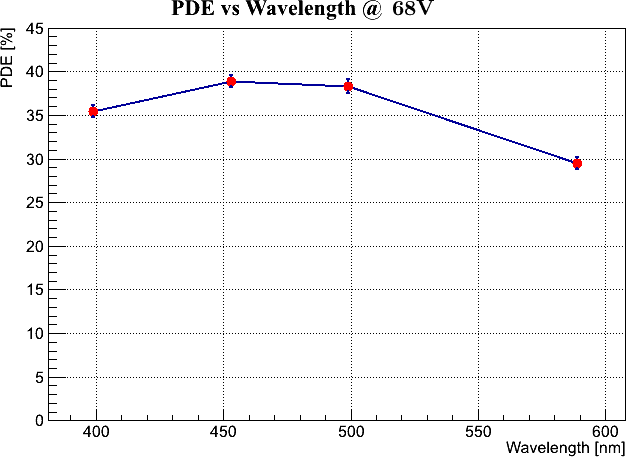}\ \hfill
      \raisebox{1cm}{
        \includegraphics[height=5cm]{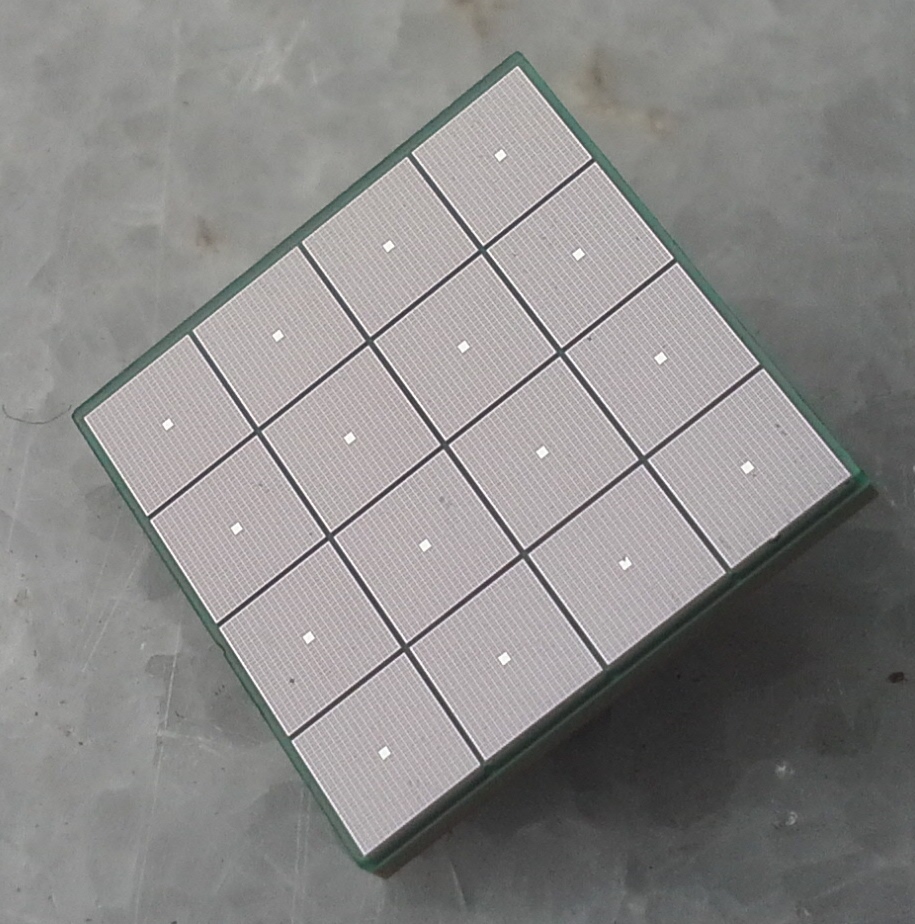}
        }
    \end{tabular}
  \end{center}
  \caption[example]{\label{fig:pde} 
    \emph{Left}: The photon detection efficiency (PDE) of the Hamamatsu S12642-0404PA-50(X) device used in the pSCT.
    \emph{Right}: Photograph of the Hamamatsu S12642-0404PA-50(X) device. The tile comprises 16 SiPMs. The through-silicon vias are visible at the very center of each SiPM.
  }
\end{figure}

The device was selected after a thorough review of available SiPMs. The primary reasons for choosing it are a peak photon detection efficiency of 40\% at 450\,nm and that it uses through-silicon vias that minimize dead space and thus increase the optical efficiency of the pSCT; see Figure \ref{fig:pde}. 

In the pSCT the SiPMs are likely biased at 3\,V above breakdown voltage. The bias of every group of four SiPMs is independently adjustable in steps of a few millivolt by a digital-to-analog converter (DAC) that is located between ground and the pixel. Drops in the bias voltage across each pixel caused by changes in the optical load (stars and background light) are compensated for by automatically readjusting the output of the DAC. The DAC features a high input impedance mode that allows turning a pixel off. The bias supply is capable of providing each pixel with a current of up to 2\,mA, allowing operation under bright moonlight conditions.

The flat fielding of the SiPM gains will be calculated using calibration events that homogeneously illuminate the camera with a time and wavelength profile similar to that of Cherenkov light. The light pulses will be generated by a flasher module consisting of 10 UV LEDs located at the center of the secondary mirror facing the camera. The flashes have a 4\,ns FWHM and a peak wavelength of 400\,nm and can be fired with an equivalent intensity between 1 and 10 LEDs to cover a large dynamic range. The brightness of each individual LED can also be adjusted by changing resistors in the driver board.

Flat fielding, gain, and linearity measurements will be done using the pulses from the LED flasher board. The photon detection efficiency (PDE) is measured in the Lab at 
five different wavelengths between 375 and 590\,nm following the method in Ref.~\citenum{2006NIMPA.567..360O}. Local muons can be used to measure the optical throughput. The nominal trigger mode will record rings produced by muons with with $E \gtrsim 6$\,GeV. A higher multiplicity, lower threshold pattern trigger can be implemented in the backplane FPGA to increase the muon trigger rate.

\subsection{Temperature stabilization of the focal plane}
\label{sec:temp}

Gain, PDE, intrinsic dark rates, and afterpulsing of SiPMs are temperature dependent. To compensate for these dependencies, the photon detectors are temperature stabilized. For that purpose the focal plane is thermally insulated, and the temperature of the photon detectors is actively stabilized with a thermoelectric element. 

Insulation is achieved with one inch thick densified SOLIMIDE foam\footnote{\url{http://www.evonikfoams.com/index.php?option=com_content&view=article&id=69&Itemid=78}} between the Delrin baseplate of the FPM and the PCBs of the quadrants and on the inside walls of the camera box that surround the focal plane. An entrance window and 50\,mm of air in between the focal plane and the entrance window insulate the focal plane from the ambient environment. For the prototype it is planned to use a 6\,mm thick window made of ACRYLITE OP4\footnote{\url{http://www.acrylite-shop.com/US/us/category.htm?$category=j09iopu9jh}}. The air inside the temperature controlled volume is dehumidified to avoid condensation on the photosensors.

\begin{figure}[t]
  \begin{center}
    \begin{tabular}{c}
      \includegraphics[height=7cm]{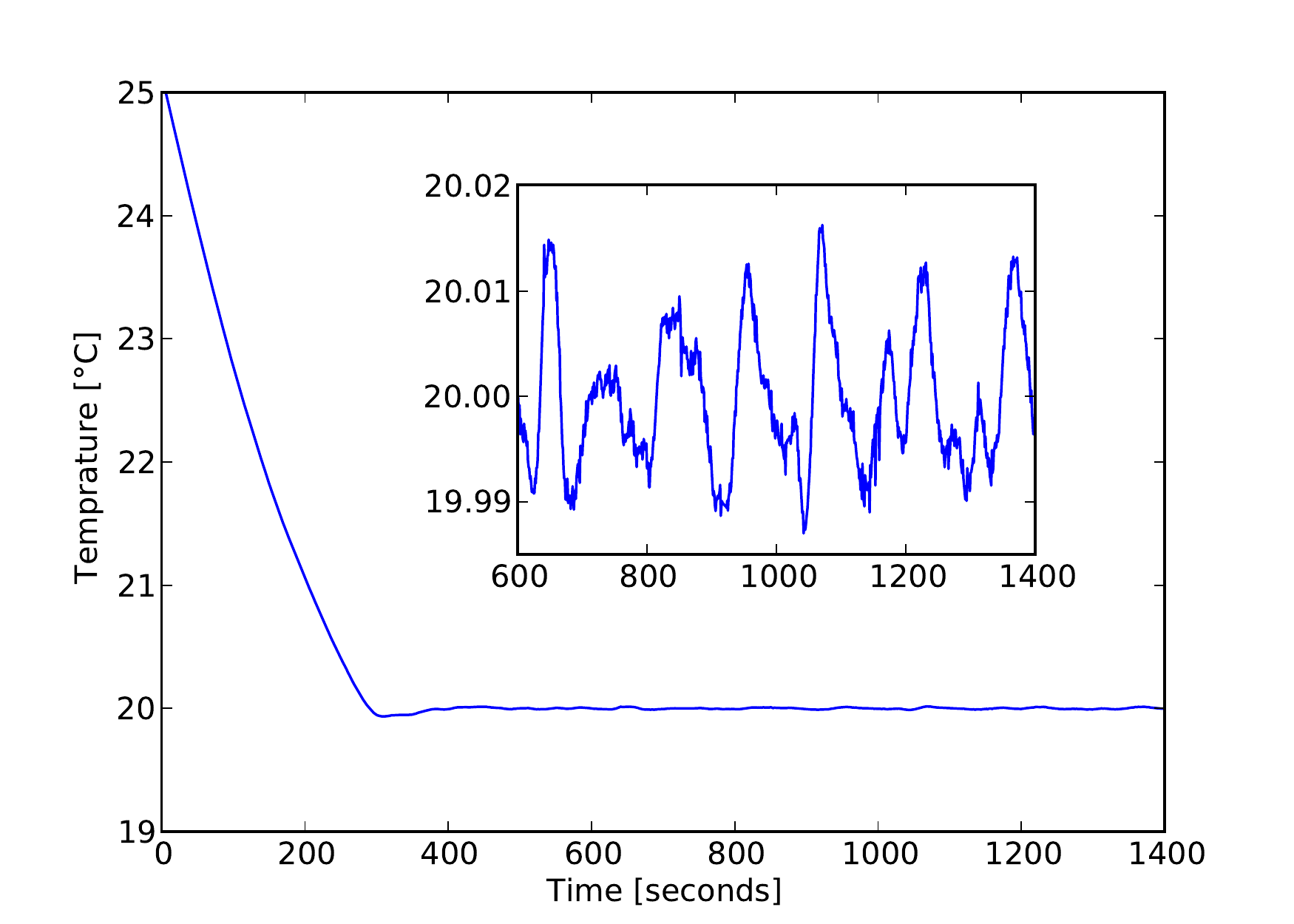}
    \end{tabular}
  \end{center}
  \caption[example]{\label{fig:TempControlTest} 
    Temperature stabilization measured on the FPM mechanical prototype. The inset shows the same data as the main plot with the vertical axis expanded.
  }
 \end{figure}

The temperature of the photon detectors is stabilized with an active control loop. The temperature regulating element is a thermoelectric element (TE) that is attached to the base of the copper post in the FPM. A heat sink is thermally coupled to the warm side of the TE and actively cooled by the fan arrays on the sides of the camera housing. The TE is controlled by an ATmega328P microcontroller programmed as  proportional-integral-derivative (PID) controller that pulse width modulates the bias of the TE. Temperature feedback is provided by thermistors mounted on the back side of the PCB of each quadrant. To improve thermal performance the PCB material is thermally conductive. Tests with prototypes show that it is possible to stabilize the temperature of the photon detectors to better than $\pm0.02^\circ$C at $20^\circ$C; see Figure \ref{fig:TempControlTest}. We estimate that it takes 1-2\,W of electrical power per module to maintain a stable temperature of the photon detectors during dark time. In observations with bright, ambient moonlight, the power will go up to 5\,W per module. 

\subsection{Front-end electronics and readout}

\begin{figure}[t]
  \begin{center}
    \begin{tabular}{c}
      \includegraphics[height=7cm]{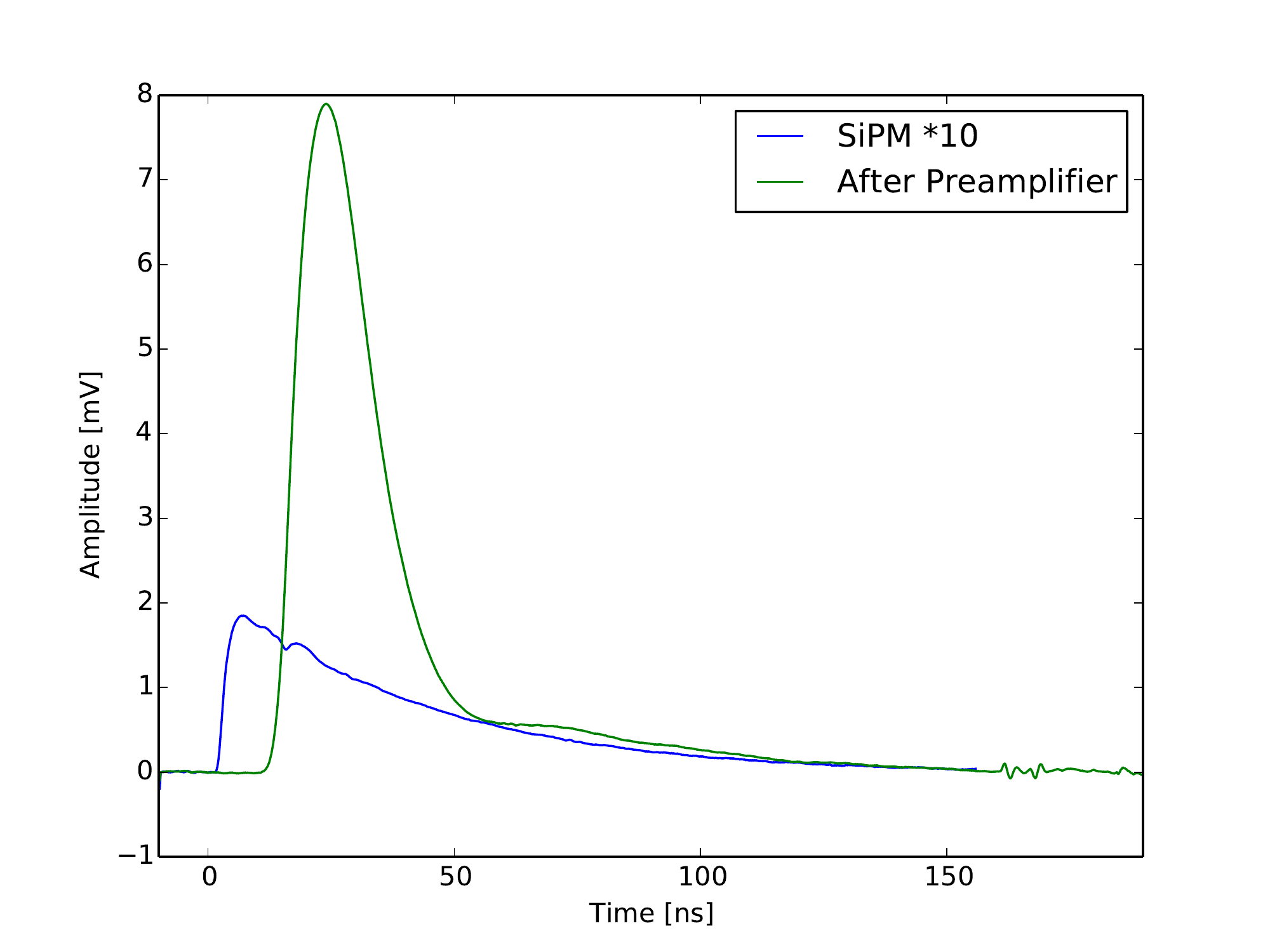}
    \end{tabular}
  \end{center}
  \caption[example]{\label{fig:waveform} 
    The waveform from the S12642-0404PA-50(X) SiPM before the preamplifier is shown in blue multiplied by a factor of 10. The signal after the preamplifier is shown in green. In the measurement the SiPM was flashed with an attenuated picosecond laser. The signal after the preamplifier is normalized to the signal produced by a single photoelectron and the signal before preamplifier normalized to the signal produced by ten photoelectrons.
  }
\end{figure}

The signals from the SiPM show a fast $\sim10$\,ns rise time followed by a slow exponential decay with a time constant in the range of $\sim300$\,ns; see Figure \ref{fig:waveform}. Before the signal is digitized and passes through the trigger electronics, it is amplified and shaped by the preamplifier. Due to the high power density of the camera, a main criterion in the design of the preamplifier is to minimize power consumption.  The final design is a two-stage amplifier that uses an AD8014 current feedback operational amplifier, which requires only 5\,mW of power per stage when supplied with 5\,volts\cite{AD8014}. The SiPM signal is differentiated by a passive, single pole high-pass filter in between the two stages. The signal at the output of the preamplifier has an amplitude of 8\,mV for a single detected photoelectron, which allows for digitization of signals with up to 250 photoelectrons per pixel within the dynamic range of the digitizer. Figure \ref{fig:waveform} shows the signal shape at the output of the preamplifier.

\begin{figure}
  \begin{center}
    \begin{tabular}{c}
      \includegraphics[height=7cm]{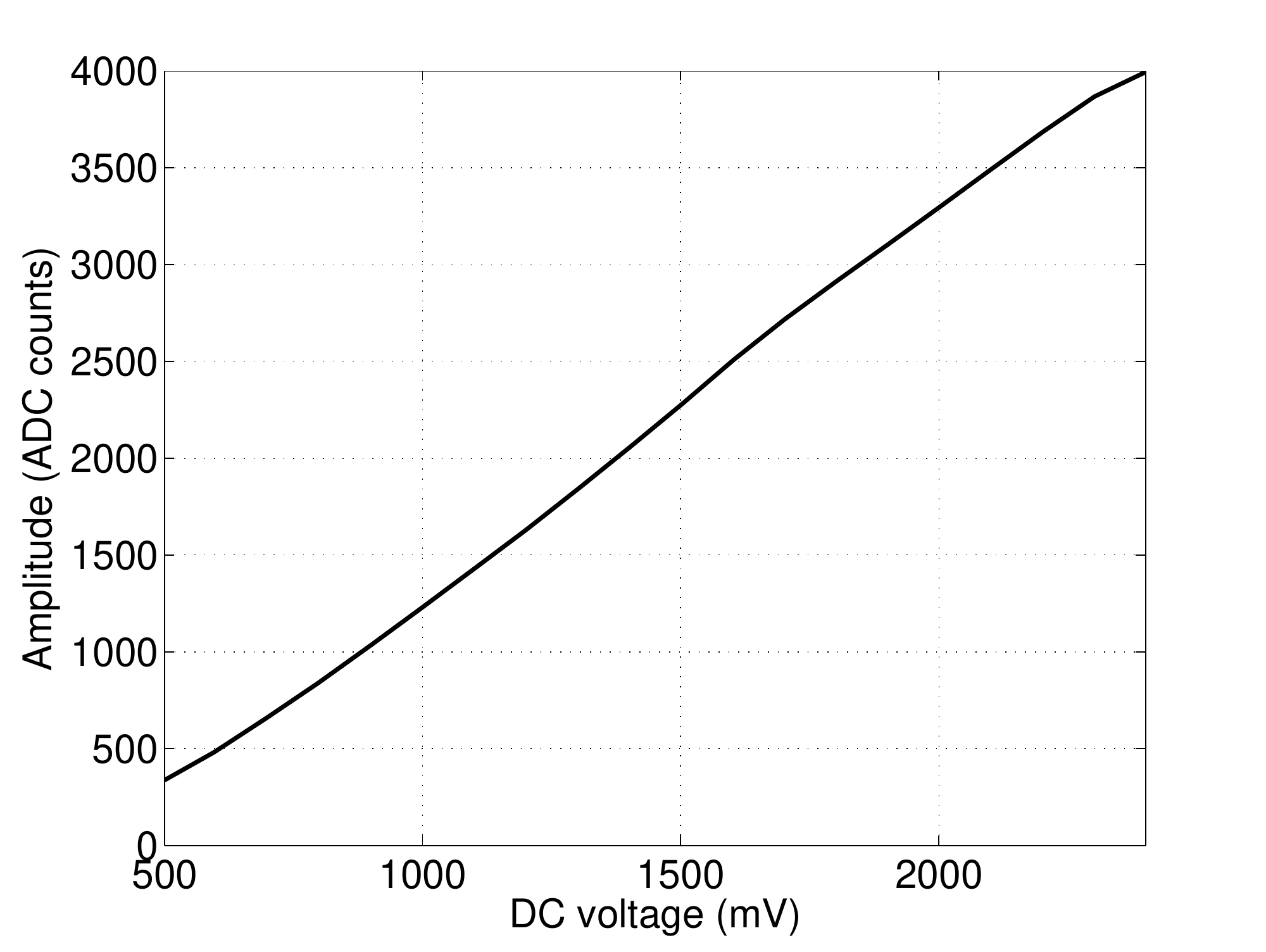}
    \end{tabular}
  \end{center}
  \caption[example]{\label{targetdynamicrange}
    Transfer function of TARGET~7 as a function of input signal amplitude.
  }
\end{figure}

After the preamplifier, the signals are digitized with TARGET~7 \cite{2012APh....36..156B}. TARGET is a cost effective solution to digitize fast signals with high channel density, deep buffering, and self triggering capability.  It is part of a family of application-specific integrated circuits originally developed for high-energy physics experiments.  The TARGET ASIC has 16 input channels per chip. Each channel has an analog ring buffer consisting of 16,384 capacitors. If a trigger occurs, the charge stored in the capacitors is sampled with an effective resolution of 10 bits by Wilkinson analog-to-digital converters. Figure \ref{targetdynamicrange} demonstrates the linearity of TARGET~7 over an input range of 1.9\,V. The linearity is better than $\pm$5\%. The intrinsic noise of TARGET~7 is $\sim$2\,mV, which enables single photoelectron resolution. The analog bandwidth of TARGET~7 is $\sim$380\,MHz.

Also included in TARGET~7 is analog summing circuitry that sums the signals from groups of four pixels. The summed signal is discriminated by a comparator with an adjustable threshold. The discriminator signal is routed to the backplane where coincidence is formed between neighboring groups of summed pixel signals. In case a coincidence is found the camera is said to have triggered. Based either on this camera trigger alone or on an array trigger requiring coincidence between neighboring telescopes, the analog pixel signals are digitized.

\begin{figure}
  \begin{center}
    \begin{tabular}{c}
      \includegraphics[height=7cm]{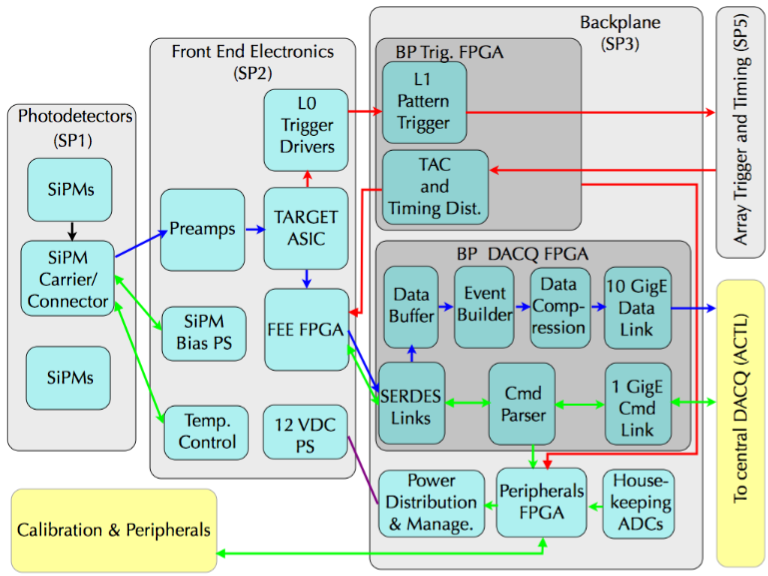}
    \end{tabular}
  \end{center}
  \caption[example]{\label{fig:example} 
    Block diagram of signal flow in the camera from the photon sensors to the backplane. 
  }
\end{figure}

An overview of the hierarchy of the signal flow in the camera is given in Figure \ref{fig:example}. The backplane synchronizes the signals between the modules, collects and processes the data stream, and distributes power to the modules. The high trigger rate expected for the SCT camera (up to about 10\,kHz) will lead to the generation of large data volumes of up to 1\,Gb/sec per camera backplane. A zero suppression scheme able to reduce this value by a factor of ten would still generate 30\,Tb/day of data. The high data rate is processed by transferring it from the backplanes to a dedicated camera server over a pair of 10\,Gbps Ethernet links in the form of UDP packets. The camera server will handle higher-level event building by combining information from multiple subfields and will also provide an interface to the CTA control system. The connection to the central data acquisition system will also enable to combining data from different telescopes by providing a clock signal that guarantees array-wide time stamp synchronization.

\section{Summary}
In this paper we presented the status of the development of a Schwarzschild-Couder telescope to boost the performance of CTA by a factor of two in its core energy region. At present, a prototype telescope is constructed that will be deployed at the VERITAS site by the end of 2015. 

The telescope uses novel two-mirror optics that results in a compact wide field camera. Novel semiconductor photon detectors are prime candidates to be used in combination with the compact camera, which is why we use SiPMs. The photon detectors are temperature stabilized to compensate fluctuations in the SiPM characteristics. Digitization of the SiPM signals takes place inside the camera with the cost effective TARGET7 chip.

\acknowledgments    
We gratefully acknowledge financial support from
the following agencies and organizations:
Ministerio de Ciencia, Tecnolog\'ia e Innovaci\'on Productiva (MinCyT),
Comisi\'on Nacional de Energ\'ia At\'omica (CNEA), Consejo Nacional
de Investigaciones Cient\'ificas y T\'ecnicas (CONICET), Argentina;
State Committee of Science of Armenia, Armenia;
Conselho Nacional de Desenvolvimento Cient\'{i}fico e Tecnol\'{o}gico (CNPq),
Funda\c{c}\~{a}o de Amparo \`{a} Pesquisa do Estado do Rio de Janeiro (FAPERJ),
Funda\c{c}\~{a}o de Amparo \`{a} Pesquisa do Estado de S\~{a}o Paulo (FAPESP), Brasil;
Croatian Science Foundation, Croatia; 
Ministry of Education, Youth and Sports, MEYS LE13012, 7AMB12AR013, Czech Republic;
Ministry of Higher Education and Research, CNRS-INSU and CNRS-IN2P3, CEA-Irfu, ANR,
Regional Council Ile de France, Labex ENIGMASS, OSUG2020 and OCEVU, France;
Max Planck Society, BMBF, DESY, Helmholtz Association, Germany;
Department of Atomic Energy, Department of Science and Technology, India;
Istituto Nazionale di Astrofisica (INAF), MIUR, Italy;
ICRR, University of Tokyo, JSPS, Japan;
Netherlands Research School for Astronomy (NOVA),
Netherlands Organization for Scientific Research (NWO), Netherlands;
The Bergen Research Foundation, Norway;
Ministry of Science and Higher Education, the National Centre for Research and 
Development and the National Science Centre, Poland;
MINECO support through the National R+D+I, CDTI funding plans and the CPAN and MultiDark
Consolider-Ingenio 2010 programme, Spain;
Swedish Research Council, Royal Swedish Academy of Sciences, Sweden;
Swiss National Science Foundation (SNSF), Ernest Boninchi Foundation, Switzerland;
Durham University, Leverhulme Trust, Liverpool University, University of Leicester,
University of Oxford, Royal Society, Science and Technologies Facilities Council, UK;
U.S. National Science Foundation, U.S. Department of Energy,
Argonne National Laboratory, Barnard College, University of California,
University of Chicago, Columbia University, Georgia Institute of Technology,
Institute for Nuclear and Particle Astrophysics (INPAC-MRPI program),
Iowa State University, Washington University McDonnell Center for the Space Sciences, USA. The research leading to these results has received funding from the
European Union's Seventh Framework Programme (FP7/2007-2013) under grant
agreement no. 262053.

\bibliography{sct}  
\bibliographystyle{spiebib}  

\end{document}